\documentclass[twocolumn]{aastex631}

\usepackage{blindtext}
\usepackage{amsmath}
\usepackage[utf8]{inputenc}
\usepackage[autostyle]{csquotes}
\usepackage{amssymb}
\usepackage{showyourwork}
\usepackage{makecell}
\let\tablenum\relaxgit 
\usepackage{siunitx}
\DeclareMathOperator*{\argmax}{arg\,max}
\DeclareMathOperator{\arcsinh}{arcsinh}

\newcommand{\chandra}{\textit{Chandra}~}
\newcommand{\chandranospace}{\textit{Chandra}}
\newcommand{\xmm}{\textit{XMM}~}
\newcommand{\fermi}{\textit{Fermi}-LAT~}
\newcommand{\ferminospace}{\textit{Fermi}-LAT}
\newcommand{\jolideco}{\textit{Jolideco}~}
\newcommand{\jolideconospace}{\textit{Jolideco}}
\newcommand{\aposteriori}{a~posteriori~}
\newcommand{\aprioir}{a~priori~}
\newcommand{\gammaray}{$\gamma$-ray\xspace}
\newcommand{\xray}{X-ray\xspace}

\begin{document}

    \title{Joint Deconvolution of Astronomical Images in the Presence of Poisson Noise}

    \author[0000-0003-4568-7005]{Axel Donath}
    \affil{Center for Astrophysics $|$ Harvard \& Smithsonian \\ Cambridge MA 02138 USA}
    \author[0000-0002-0905-7375]{Aneta Siemiginowska}
    \affil{Center for Astrophysics $|$ Harvard \& Smithsonian \\ Cambridge MA 02138 USA}
    \author[0000-0002-3869-7996]{Vinay L.\ Kashyap} 
    \affil{Center for Astrophysics $|$ Harvard \& Smithsonian \\ Cambridge MA 02138 USA}
    \author[0000-0002-0816-331X]{David A.\ van Dyk}
    \affil{Department of Mathematics, Imperial College London \\ London SW7 2AZ UK}
    \author[0000-0003-4428-7835]{Douglas Burke}
    \affil{Center for Astrophysics $|$ Harvard \& Smithsonian \\ Cambridge MA 02138 USA}
    \correspondingauthor{Axel Donath}
    \email{axel.donath@cfa.harvard.edu}

    \begin{abstract}
    
        We present a new method for joint likelihood deconvolution (\jolideconospace) of a set of astronomical observations of the same sky region in the presence of Poisson noise. 
        The observations may be obtained from different instruments with  different resolution, and different point spread functions. \jolideco reconstructs a single flux image by optimizing the posterior distribution based on the joint Poisson likelihood of all observations under a patch-based image prior. The patch prior is parameterised via a Gaussian Mixture model which we train on high-signal-to-noise astronomical images, including data from the  James Webb Telescope and the GLEAM radio survey. This prior favors correlation structures among the reconstructed pixel intensities that are characteristic of those observed in the training images. It is, however, not informative for the mean or scale of the reconstruction. By applying the method to simulated data we show that the combination of multiple observations and the patch-based prior leads to much improved reconstruction quality in many different source scenarios and signal to noise regimes. We demonstrate that with the patch prior \jolideco yields superior reconstruction quality relative to alternative standard methods such as the Richardson-Lucy method. We illustrate the results of \jolideco\ applied to example data from the \chandra X-ray Observatory and the \ferminospace\ Gamma-ray Space Telescope. By comparing the measured width of a counts based and the corresponding \jolideco flux profile of an X-ray filament in SNR {\it 1E 0102.2-7219} we find the deconvolved width of $0.58\pm 0.02$ arcsec to be consistent with the theoretical expectation derived from the known width of the PSF.
    \end{abstract}

    \section{Introduction}
\begin{figure*}
        \begin{centering}
            \includegraphics[width=\linewidth]{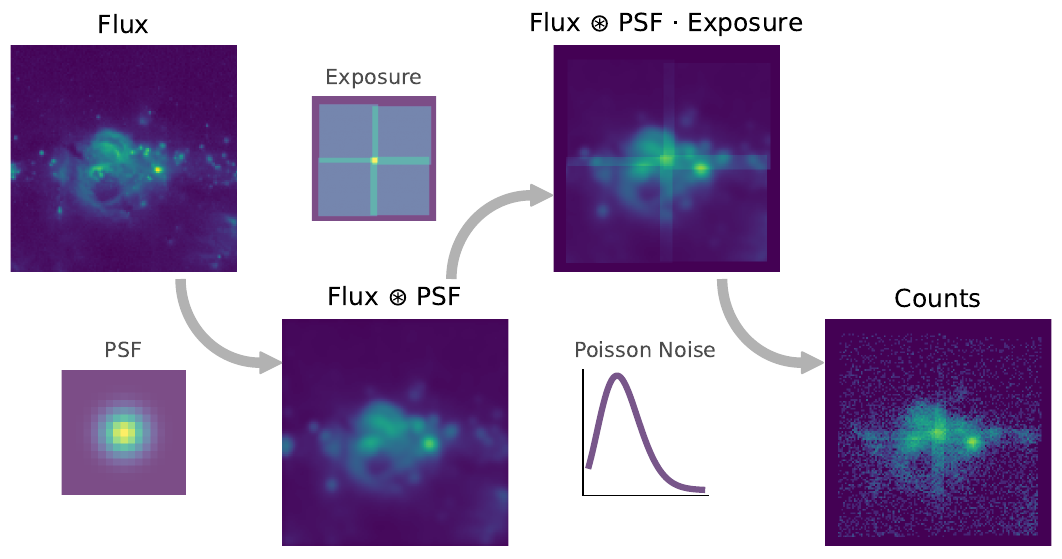}
            \caption{
                Illustration of the forward model of the low-count imaging process. The flux represents the true unknown flux distribution in the sky. The flux is convolved with an image of the PSF, which represents an estimate of the resolution of the imaging instrument. The exposure consists of the superposition of multiple independent observations with different observation time and pointing position. The exposure weighted flux image is degraded by Poisson noise.
            }
            \label{fig:model-illustration}
            \script{model-illustration.py}
        \end{centering}
    \end{figure*}

    The resolution of any physical or astronomical imaging process is inherently limited by the instrument or telescope used. Furthermore, the quality of the resulting image is affected by background, instrumental measurement noise, and non-uniform exposure. This is especially true in the low signal-to-noise regime, where images are affected by Poisson noise. This includes images taken by \xray and \gammaray~telescopes but also medical images obtained with methods such as positron emission tomography or single-photon emission computed tomography. This low-count imaging process is illustrated in Figure~\ref{fig:model-illustration}.

    To maximise the scientific use of available data it is desirable to correct for the instrumental imprint on the data as well as reduce its noise level. The process of improving the resolution of an image in post-processing is often called \enquote{deblurring} or \enquote{deconvolution}, as it aims to invert the effect of the data being degraded by a forward convolution through the point spread function (PSF) of the instrument. In general this class of problems is categorised as \enquote{inverse problems}. If there is prior information available on the shape of the PSF, we typically refer to the process more precisely as \enquote{unblind deconvolution}. In case of astronomical observations the PSF can for example be estimated empirically from observation of known point sources, such as distant stars, or a model of the PSF can be obtained from simulation of the telescope or instrument.

    \subsection{Related Work}
    \label{sec:related}
    In the literature there have been many efforts to solve the unblind deconvolution problem for images affected by Poisson noise. Most methods rely fundamentally on modeling image generating as a linear process. Given a known PSF a prediction of the discrete measured image can be obtained via \enquote{forward folding} and be compared against an observation assuming Poisson statistics in each pixel. However, the methods differ in their additional prior or regularisation assumptions as well as their approach to optimization or sampling of the underlying likelihood function or posterior distribution. In the following section we give a short overview over these methods.
    
    The standard baseline method for image deconvolution in the presence of Poisson noise was proposed independently by \cite{Richardson1972} and \cite{Lucy1974}. (We refer to this method as \enquote{RL}.) The method involves estimating the reconstructed image by an iterative procedure that converges to a maximum likelihood solution of the problem.
    
    However, in practice the RL method has multiple, partly related limitations: first, when the number of iterations is too high, the method shows a tendency to decompose regions of constant brightness into a set of bright emission peaks (\enquote{speckles}). This problem was described and studied, for example,  by \cite{Reeves1995} and \cite{Fish1995}. This means that in practice the RL method is not iterated until it converges, leading to the second limitation: there is no objective criterion for stopping the optimization process. Usually the iteration is stopped when the reconstructed image seems plausible to the eye. 
    Lastly, the RL method only provided a \enquote{point estimate} and no associated estimate of the uncertainties of the reconstruction. Uncertainties are desirable especially in the Poisson domain, for example, to measure the significance of specific features found in the reconstructed image.
    
    To address the issue with RL decomposing extended emission structures into groups of point sources, multiple additions to the RL method  have been proposed. Most importantly this lead to the introduction of regularised RL methods, such as use of total variation \citep{Dey2006} or sparse representations based on \enquote{wavelets} or \enquote{curvelets} \citep{Starck2003}. The challenge is rooted in two fundamentally contradicting goals: a deconvolution process that favors point sources will decompose extend structures present in the image, while introducing local correlations between neighbouring pixels will lead to less sharp reconstruction of point sources.

    One possible approach to overcome this challenge is proposed by \cite{Esch2004} and \cite{Connors2011} with the EMC2 and LIRA methods. Their key idea is to introduce a prior assumption that favors smoothness on multiple spatial scales of the image. Instead of a trying to find a single point estimate for the reconstructed image they also sampled from the posterior distribution, which allowed them to quantify the associated uncertainties. However, the prior distribution is based on the generic assumption of smoothness and does not encode any \enquote{real world} or physical information on the structures present in the image.

    An interesting modification of the standard RL method is proposed by \cite{Ingaramo2014}. They show that RL could be effective in creating a merged image from multiple observations. They also observed that RL would take the best visual properties out of each image and merged them to result in a reconstruction closer to the \enquote{ground truth}.
    
    \cite{Selig2015} and \cite{Pumpe2018} approach the problem of deconvolution of Poisson noise data implicitly using the framework of information field theory. Based on physical prior assumptions on the spatial distribution of point sources and local correlation structures of diffuse emission, they are able to deconvolve \fermi observations
    into a mixture of point-like and diffuse emission components. While their method works well for large-scale diffuse emission, such as the Galactic diffuse gamma-ray emission, it shows artifacts for smaller extended emission structures such as morphological details of extended sources in the Galactic Plane.
    
    Other interesting approaches to the problem of deconvolution can of course be found in the field of computer vision. For example \cite{Zoran2011} build on very similar assumption of local correlation for natural images. However, instead of assuming a fixed local correlation structure, such as smoothness defined by a Gaussian kernel with a given bandwidth determining the correlation scale, they proposed to learn the local correlation structure on millions of patches extracted from a set of reference images. For this they propose to split training images from representative datasets into small patches and learn a  Gaussian Mixture Model (GMM) for the distribution of extracted patches. To use the GMM in reconstruction they introduce the concept of expected patch likelihood (EPLL). They show that the patch-based GMM prior leads to much improved image reconstructions for multiple inverse problems such as denoising, deblurring and inpainting. \cite{Bouman2016} later adapted the patch prior successfully for the reconstruction of sub-millimeter radio data from the Event Horizon Telescope (EHT).
    
    \subsection{Some Remarks on Deep Learning Methods}
    In the past decade Machine learning (ML) especially deep convolutional neural networks (CNN) became the standard for deblurring and super-resolution of natural images \citep{Zhang2022}. However, their success relies on the availability of massive sets of training data with typically low noise and corresponding ground truth to perform supervised learning. For a variety of reasons, which we outline in the following, this approach is not ideal for astronomical \xray and \gammaray data.
    
    Firstly, astronomical \xray and \gammaray data are rare, expensive to obtain, and there is no ground truth, because we can only observe the sky with our given instruments. To circumvent this issue one could rely on transfer learning and simulated data. For example, radio or optical images could be forward folded with a known PSF and degraded with Poisson noise to generate training data.  Astronomical images, however, have a large dynamic range and show a high diversity in morphological structure, which would requires a large amount of training data.  

    Nonetheless, this approach has been successfully applied for specific classes of sources, such as images of Galaxies in the Sloan Digital Sky Survey (SDSS) \citep{Schawinski2017} or as a post processing step of results obtained with regularised RL \citep{Akhaury2022}. Recently, \cite{Sweere2022} used images from the large scale structure simulations with promising results.

    In addition to the challenge of obtaining training data, it is known that the PSF of instruments such as \chandra is highly variable. It varies with observation conditions and depends, for example, on energy as well as offset angle from the pointing direction. To cover the whole parameter space of the PSF one would need to train a network for each of the PSF models.

    Some of the challenges of deblurring with CNNs are due to the convolutional nature of the networks. In a single layer, linear approximation, the network would need to learn the inverse filter of the PSF, corresponding to the well known \enquote{Wiener Filter} solution of the problem. This requires either large convolution kernels or deep network architectures, which in turn require large amounts of training data. This challenge was described early on by \cite{Li2014} who introduced a baseline architecture for a deep CNN for image deblurring.
    
    One last important challenge for deep learning based methods is obtaining uncertainty estimates for the reconstructed image from CNNs. In the limit of Poisson noise it is desirable to trace the full posterior distribution
    which is not possible for most network architectures, with the exception of, for example, normalising flows. This problem could be addressed with Monte Carlo (MC) based methods, such as bootstrapping, which further increases the demand on training data and computational resources. For these reasons we take a statistical modelling approach to the deconvolution problem in the presence of Poisson noise. We do, however, include elements from ML to represent the prior information on images.

    \section{Method}
    
    \subsection{Poisson Joint Likelihood}
    Our goal is to recover an image $\mathbf{x}$ from a set of multiple low-count observations. Most generally we assume our total dataset consists of $J$ individual observations of the same region of the sky, which we jointly model. The assumption is that the underlying, unknown \enquote{true emission} image, we are interested in, does not change over the timescale of the observations. Under this assumption the datasets can be for example:

    \begin{itemize}
        \item Different observations with one instrument or telescope at different times and observation conditions. For example, multiple observations of \chandra of the same astrophysical object with different offset angles and exposure times;
        \item Observations from different telescopes, which operate in the same wavelength range. For example, a \chandra and \xmm observation of the same region in the sky;
        \item A single observation with one telescope with different data quality categories and different associated instrument response functions, such as event classes for \fermi.
    \end{itemize}

    Any arbitrary combination of the possibilities is also possible.
    In principle, we can combine observations in different energy ranges. In this case, however, the assumption of the morphology not changing with energy would apply, which is rarely fulfilled for astronomical objects.
    
    For each individual observation $j$ the predicted counts can be modelled by forward folding the unknown flux image $\mathbf{x}$ through the individual instrument response:
    \begin{equation}
        \label{eq:model}
        \boldsymbol{\lambda}_j(\mathbf{x}) = \mathbf{E}_j \cdot \left(\mathrm{PSF}_j \circledast \mathbf{x}\right) + \mathbf{B}_j,
    \end{equation}
    where the expected counts $\boldsymbol{\lambda_j}$ are given by the convolution of the unknown flux image $\mathbf{x}$ with the observation specific point spread function $\mathrm{PSF}_j$ and an observation specific image of the exposure $\mathbf{E}_j$ and background emission $\mathbf{B_j}$ can be optionally taken into account. A visual representation of the model of the imaging process for a single observation is shown in Figure~\ref{fig:model-illustration}.

    Given a single observation $j$ of an unknown flux image $\mathbf{x}$ and assuming the noise in each pixel $i$ in the recorded counts image $\mathbf{D}_j$ follow independent Poisson distributions with expectation $\boldsymbol{\lambda}_j$, the likelihood $P$ of obtaining the measured image with $N$ pixels is given by:
    \begin{equation}
        \label{eq:poisson}
        P\left( \mathbf{D}_j | \boldsymbol{\lambda}_j \right) = \prod_{i=1}^N \frac{{e^{ - \lambda_{j,i} } \lambda_{i,j} ^ {D_{j,i}}}}{{D_{j,i}!}},
    \end{equation}
    where $\boldsymbol{\lambda}_j$ depends on $\mathbf{x}$ as defined by Equation~\ref{eq:model}. By taking the logarithm, multiplying by $-1$, and dropping the constant terms one can transform the product into a sum over pixels, which is also often called the \enquote{Cash} \citep{Cash1979} fit statistics:
    \begin{equation}
        \label{eq:cash}
        \mathcal{C}\left( \mathbf{D}_j | \boldsymbol{\lambda}_j \right) = \sum_{i=1}^{N_j} \left(\lambda_{j,i} - D_{j, i} \log{\lambda_{j,i}}\right),
    \end{equation}
    By summing over all observations we obtain the joint log-likelihood of measuring a set of counts images $\mathbf{D}$ given an unknown flux image $\mathbf{x}$:
    \begin{equation}
        \label{eq:joint}
        \mathcal{L}\left( \mathbf{D} | \mathbf{x} \right) = \sum_{j=1}^J \mathcal{C}\left( \mathbf{D}_j | \mathbf{x} \right)
    \end{equation}
    In principle, we could obtain the Maximum Likelihood Estimate (MLE), $\hat{\mathbf{x}}$ of ${\mathbf{x}}$, by directly optimizing Equation~\ref{eq:joint}. Indeed, this is the objective of the RL algorithm.  However, in general this represents an \enquote{ill posed} inverse problem. Estimates for $\mathbf{x}$ which have similar likelihood values $\mathcal{L}(\hat{\mathbf{x}}_1)$ may look very different from each other and might also look different compared to a given ground truth, if available. Both the convolution operation
    and the noise lead to a loss of information in the data which cannot easily be recovered. The likelihood surface shows many local, close-by maxima. Depending on the starting value used for $\mathbf{x}$ and the optimization algorithm the computed MLE is likely to be a local maxima. This is fundamentally the reason for RL decomposing estimates into point sources, described in Section~\ref{sec:related}.
    
    \subsection{A Posteriori Distribution}
    When reconstructing images in the context of inverse problems we are operating in a high dimensional parameter space, because each pixel in the image represents a parameter in the optimization process. By using prior information we can introduce correlations among the parameters, to reduce the \enquote{effective dimensionality} of the problem and guide the optimization process toward unique and more stable solutions.
    
    Using Bayes' rule we can express the \aposteriori likelihood under a given prior distribution, $P(\mathbf{x} )$:
    \begin{equation}
        \label{eq:bayes}
        P(\mathbf{x}|\textbf{D}) = P(\mathbf{x} ) \frac{P(\textbf{D} |\mathbf{x})}{P(\textbf{D})}.
    \end{equation}
    The Maximum A Posteriori (MAP) estimate of $\mathbf{x}$ is the value that maximizes Equation~\ref{eq:bayes}.
    Taking the logarithm of Equation~\ref{eq:bayes}, replacing the definition of the likelihood $P(\mathbf{D}|\mathbf{x})$ with Equation~\ref{eq:joint} and dropping the normalization term $P(\mathbf{D})$ which is independent of $\mathbf{x}$, we obtain the following expression for the log-posterior distribution
    $\mathcal{L}$:
    \begin{equation}
        \label{eq:total}
        \mathcal{L}\left(\mathbf{x} | \mathbf{D} \right) = \sum_{j=1}^J \mathcal{C}\left( \mathbf{D}_j | \mathbf{x} \right) - \beta \cdot \mathcal{P}(\mathbf{x}),
    \end{equation}
    where $\mathcal{C}\left( \mathbf{D}_j | \mathbf{x} \right)$ represents the summed log-likelihood for an individual observation $j$ and $\mathcal{P}(\mathbf{x})$ represents the log of the prior distribution with the factor $\beta$ adjusting the weight of the prior relative to the joint log likelihood term. The parameter $\beta$ can be viewed as a hyper-parameter, chosen by the user to adjust the strength of the prior distribution relative to the likelihood. The choice of $\beta$ is optional and for all experiments presented in this paper we set $\beta=1$. 

    \subsection{Patch-based Image Prior}
    \label{sec:patch-prior}
    In general it is difficult to capture global image characteristics in a prior distribution taking into account all pixel to pixel correlations. The size of the corresponding correlation matrix would grow with the squre of the number of pixels  and thus quickly become computationally untraceable. However, astronomical images typically only exhibit local correlations among the pixels. This is because the spatial scale of astrophysical processes in an image is always limited by the finite interaction time and the fact that the interaction cannot propagate faster than the speed of light. This means on smaller scales astronomical images often contain basic structures such as edges, corners, filaments or periodic patterns formed by local astrophysical processes. For this reason we adopt the patch-based image prior distribution introduced by \cite{Zoran2011}. 

    A patch is defined as a small, typically square, region of size $N_p \times N_p$ pixels, representing a local neighbourhood in a given image. Mathematically we can formulate the extraction of such patches as:
    \begin{equation}
        \mathbf{x}_n = \mathbf{P}_n \mathbf{x},
    \end{equation}
    where $\mathbf{P}_n$ represents a neighbourhood matrix to extract the $n$-th patch from the image $\mathbf{x}$ and where $\mathbf{x}_n$ is the vector of length $N_p^2$ representing the \enquote{flattened} image patch. By obtaining a large number of patches we can model the distribution of pixel intensities within the patches using a Gaussian Mixture Model (GMM). The GMM is a universal function approximator and thus can model arbitrary distributions. For now we assume that we have a pre-trained GMM which encodes the patch distribution of some reference image we have chosen. During the actual image reconstruction we aim to identify the most likely GMM component per patch, treating this as a discrete nuisance-parameter that is optimized along with the image $\mathbf{x}$. 
    
    A convenient property of the GMM is that its probability density can be evaluated in closed form. For a given single patch $\mathbf{x}_n$,
    \begin{equation}
        \label{eq:gmm}
        P_{\rm GMM}(\mathbf{x}_n) = \sum_{k=1}^K \pi_k \mathcal{N}(\mathbf{x}_n| \boldsymbol{\mu}_k, \boldsymbol{\Sigma}_k^2),
    \end{equation}

    Where $\mathcal{N}$ is the Gaussian normal distribution. If we knew the relevant component, $k_n$, for patch $\mathbf{x}_n$ we could derive the conditional log-prior distribution directly from Equation~\ref{eq:gmm}:
    \begin{equation}
    \begin{split}
    \mathcal{P}_{\rm GMM}(\mathbf{x}_n|k_n) = -\frac{1}{2} \left[ \right. d \log(2\pi)
    + \log(|\boldsymbol \Sigma_{k_n}|)\\
    + (\mathbf{x}_n - \boldsymbol{\mu}_{k_n})^T \Sigma^{-1}_{k}(\mathbf{x}_n - \boldsymbol{\mu}_{k_n}) \left. \right].
    \label{eq:cond.log.prior}
    \end{split}
    \end{equation}
    If we split the image $\mathbf{x}$ into multiple non-overlapping patches, for each patch $\mathbf{x}_n$, we can find the GMM component $\hat{k}_n$, which maximises the conditional log-prior distribution given in Equation~\ref{eq:cond.log.prior}:
    \begin{equation}
        \hat{k}_n = \underset{k \in \{1, ..., K\} }{\argmax}{\mathcal{P}_{\rm GMM}(\mathbf{x}_n|k)}
        \label{eq:khat}
    \end{equation}
    \vspace{0.2em}
    As the number of components in the GMM is typically limited to $\mathcal{O}(10^2)$ we can optimize the conditional log-prior distribution by simply evaluating it for each of the GMM components and choose the one with the maximum value. Finally, to evaluate the (optimized) log-prior of the full image we sum the maximum values of the individual patches: 
    \begin{equation}
    \label{eq:prior-total}
        \mathcal{P}(\mathbf{x}, \hat{\mathbf{k}}) = 
        \sum_{n = 1}^N \mathcal{P}_{\rm GMM}(\mathbf{P}_n \mathbf{x} | \hat{k}_n) +
        \sum_{n = 1}^N \log \pi_{\hat{k}_n},
    \end{equation}   
    where $\mathbf{P}_n$ is again the matrix that extracts the $n$-th patch from the image $\mathbf{x}$ to be reconstructed.
    
\begin{figure}
        \begin{centering}
            \includegraphics[width=\linewidth]{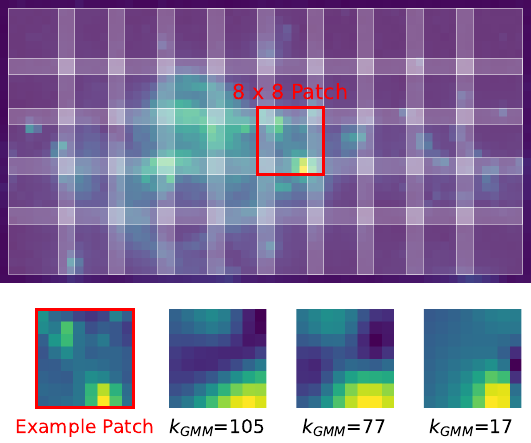}
            \caption{
                Grid of overlapping quadratic patches of size $8\times8$~pixels on top of an example image of the size $128 \times 68$~pixels. The overlap size was chosen to be $2$ pixels, which results in 50 patches in total. The patches are shown translucent, such that the horizontal and vertical \enquote{stripes} result from two neighboring overlapping patches, while the small, even lighter squares result from four patches overlapping at the their corners. The bottom row shows the red example patch $\mathbf{x}_n$ on the left. The three other patches show the example patch represented by a linear combination of the six largest Eigenvectors of the covariance matrices of the 3 most likely GMM components $k = 105, 77, 17$. This example uses the GMM learned from the JWST Cas A image.
            }
            \label{fig:patches}
            \script{patches.py}
        \end{centering}
    \end{figure}

    The bottom row of Figure~\ref{fig:patches} shows an example of a patch along with its representation in the \enquote{Eigenbasis} of the covariance matrices of its three most likely GMM components. Visually, the representation of the patch in the Eigenbasis of the covariance results in a denoised version of the patch.
    
    \cite{Zoran2011} found that the reconstruction quality can be improved by working with a grid of overlapping patches. This is illustrated in Figure~\ref{fig:patches}. However, using a grid of overlapping patches makes the interpretation of the prior distribution more difficult, because pixels at the corners and edges of patches are counted multiple times. For this reason we introduce a weighting mask when evaluating the GMM prior distribution. This mask weighs the pixels of each patch by their overlap with neighboring patches. Overlapping edges are assigned a weight of 1/2, while overlapping corners are assigned a weight of 1/4, when computing the log prior distribution for a patch as given by Equation~\ref{eq:cond.log.prior}.
    
    Apart from this, the process for choosing the most likely GMM prior component remains unmodified. However, we note that we neglected the prior dependence between neighbouring patches\footnote{That is, we continue to evaluate the log-prior distribution of the full image by summing over the overlapping patches as in Equation~\ref{eq:prior-total}, even though overlapping patches include common pixels and are thus clearly not \aprioir independent.}. As the choice of the GMM component per patch is discrete, a simultaneous gradient based optimization of all patches is not easily possible, because we cannot compute gradients across GMM components. This would require a continuous representation or interpolation between GMM components.
   
    \subsubsection{Image Dynamic Range and Patch Normalization}
    \label{ssec:patch-norm}
    Astronomical images typically show a much higher dynamic range compared to natural images. To normalize the patches \cite{Zoran2011} proposed to subtract the patch mean before learning the GMM. This leads to the GMM modeling the same morphological features at different levels of contrast as different components. While this proved to be beneficial in the context of reconstructing natural images it does not scale to images with high dynamic range, because every morphological feature would need to be present at many levels of intensity. While this could be handled by introducing significantly more components to the GMM, it seems unfeasible in practice due to the limited availability of training data and computational resources. For this reason we instead propose a modified normalization for the patches:
    \begin{equation}
        \label{eq:patch-norm}
        \tilde{\mathbf{x}}_n = \frac{\mathbf{x}_n - \Bar{\mathbf{x}}_n}{\Bar{\mathbf{x}}_n}
    \end{equation}
    By dividing by the mean of the patch, $\Bar{\mathbf{x}}_n$, the structure in the patch becomes independent of the total brightness the patch contains. This results in patches being grouped only by their pixel to pixel contrast and not by their overall brightness. This corresponds to the assumption that the structure in the patch is \aprioir independent of the total flux contained in the patch, which we consider a valid assumption for astronomical images. This also generalizes to the prior distribution of the whole image described by Equation~\ref{eq:prior-total}. The value of the prior distribution is independent of the total flux of the image. The prior distribution simply regularizes the structure in the image, while the overall intensities are typically well constrained by data as quantified by the Poisson likelihood term. Thus we simply assume a uniform prior distribution for the patch intensities. Algorithmically, we replace $\mathbf{x}_n$ with $\tilde{\mathbf{x}}_n$ when evaluating Equations~\ref{eq:gmm} and \ref{eq:khat} and transform $\mathbf{P}_n\mathbf{x}$ via Equation~\ref{eq:patch-norm} when evaluating Equation~\ref{eq:prior-total}.

    \subsubsection{Cycle Spinning}
    The evaluation of the patch prior divides the reconstructed image into a fixed grid of overlapping patches. To avoid artifacts in the reconstruction due to the fixed position of the grid, we use the method of \enquote{cycle spinning} \citep{Coifman1995}. Before evaluating the patch prior at each iteration of our optimizer, we shift the image $\mathbf{x}$ by a random number of pixels in both the x and y direction, where the shift is less than or equal to half the size of a patch. The same method has, for example, been used in \cite{Esch2004}.
    
    A variation of cycle spinning specifically for the patch prior is proposed in \cite{Parameswaran2018}. Instead of a regular grid for the patches they propose to work with a randomised grid of patches, where the center of each patch is randomly shifted, while ensuring that patches overlap by at least one pixel. In doing so they find that a smaller total number of patches can be used, while retaining the same reconstruction quality. This improves the computational performance of the method. We also implemented the evaluation of the patch prior on a randomized patch grid for comparison and found similar results as \cite{Parameswaran2018}: the reconstruction is slightly faster with the same reconstruction quality. However for all results shown later in the manuscript we chose the simpler cycle spinning described in the previous paragraph.

    \subsection{Systematic Errors of Predicted Counts}
    Each observation $\mathbf{D}_j$ is subject to systematic errors. For example, the pointing accuracy of the telescope when repeatedly pointed to the same object or when using different types of telescopes for observation, the reconstructed position of an astronomical object may vary among the observations. The process of correcting for these inaccuracies is called \textit{astrometry} and is a standard procedure in for example \chandra data analysis. A second example of systematic errors is the variation of background emission between multiple observations.
    
    To account for both systematic effects we extend the model in Equation~\ref{eq:model} for the predicted counts, by a background normalization factor $\alpha_j$ and a linear shift in position $\phi_j(\mathbf{x}| \delta_{x,j}, \delta_{y,j})$, which is specific for each observation. This yields the following modified model:
    \begin{equation}
        \label{eq:model-npred-calibration}
        \mathbf{\lambda}_j = \mathbf{E}_j \cdot  \left(\mathrm{PSF}_j \circledast \phi_j(\mathbf{x}| \delta_{x,j}, \delta_{y,j})\right) + \alpha_j \cdot \mathbf{B}_j
    \end{equation}
    In practice one observation is used as a reference and its corresponding systematic shifts in position $\delta_{x,j}, \delta_{y,j}$ are frozen during optimization. An example of the effect of the positional shift and background normalization can be seen in Figure~\ref{fig:example-fermi-lat-non-calibrated}.

    \subsection{Estimation of Statistical Errors}
    Estimating the global covariance matrix for estimated image is computationally infeasible, because the size of the matrix grows with the squared number of pixels. However, there are other ways to estimate the statistical uncertainties associated with the flux image. We propose two methods:

    \begin{itemize}
    \item \textbf{Bootstrapping Observations}. Given that multiple observations of the same region of the sky are available, \jolideco can be run repeatedly on a random subset of images. This results in a set of reconstructed images representing a distribution from which we can estimate the errors on $\mathbf{x}$. Depending on the size and number of the images the associated computational effort can be very large as it requires running \jolideco many times to sufficiently describe uncertainty in the fitted image.

    \item \textbf{Bootstrapping Events}. If only one or a small number of observations are available we propose bootstrapping of the events used to compute the counts image(s). This corresponds to splitting a single observation into multiple observations by splitting the single exposure into multiple time intervals. This way one can apply the first proposed method.
    \end{itemize}

    We also experimented with computing the diagonal of the Hessian of the posterior distribution evaluated at the MAP estimate. However we found this numerically challenging because of the random cycle spinning. 

    \section{Implementation}
    \subsection{\jolideco Framework}
    The goal of the reconstruction process is to optimize the a-posteriori likelihood defined by Equation~\ref{eq:total}. Given that there is a parameter for each pixel in the image, this is a high dimensional optimization problem. Differentiable programming frameworks such as PyTorch~\citep{Pytorch2019} allow for solving high dimensional optimization problems using back-propagation. They also offer a variety of optimization methods such as stochastic gradient decent (SGD). The latter is also applicable in our case, as we work with multiple observations at the same time, where a subset of the observations represents a \enquote{mini batch} in terms of ML.

    We implemented the \jolideco method as an independent Python package, based on PyTorch as the computational back-end.  We defined a modular, object oriented code structure, to allow parts of the algorithm to be flexible and interchangeable, such as the choice of the patch normalization, the choice of GMM, optimization methods, and serialisation formats for the reconstruction results as well as the corresponding diagnostic information, such as the trace of the posterior and prior likelihood values. This will simplify future improvement and extension of the package as well as the adoption of \jolideco by the astronomical community. The package is available on the platform GitHub\footnote{\url{https://github.com/jolideco/jolideco}}.

    We use Numpy~\citep{Numpy2020} for handling data arrays and Matplotlib~\citep{Hunter2007} for plotting. To learn the GMM from patches we use the standard implementation provided in the Scikit-Learn~\citep{scikit-learn} package. To handle the serialisation of images to the FITS data format and to handle world coordinate systems (WCS) we use Astropy~\citep{Astropy2018}.

    \subsection{Learning GMMs}
    \label{ssec:jolideco-gmm-library}
    The patch-based image prior described in Section~\ref{sec:patch-prior} requires a pre-trained GMM. We train the GMM by choosing a set of reference images, splitting them into patches of size $8\times8$ pixels and learning a GMM with $K$ components from the whole distribution of patches. In general the strength of the GMM prior is its high \enquote{expressiveness}. This means the prior distribution contains a large variety of structural shapes and has freedom to adapt to the data in the optimization process by choosing the most likely GMM component for each patch in the image. With radio data, \cite{Bouman2016} found their results to be mostly independent of the choice of data used for learning the GMM prior. 
    
    However, in the low signal-to-noise regime we typically expect a higher dependence of the results on the choice of prior, simply because the prior distribution is more influential when the data is less informative. Thus, there is the potential for the choice of GMM training data to be influential on results and the prior should be adapted to the particular analysis scenario in order to avoid bias. For this reason and for purposes of testing we learned and provide a selection of GMMs to be used with the patch prior for download in a \textit{GitHub} repository\footnote{\url{https://github.com/jolideco/jolideco-gmm-library}}. 
    
    \subsubsection{Zoran \& Weiss}
    As a baseline reference we use the original trained GMM provided by \cite{Zoran2011}. It was trained from the Berkeley image database \citep{Martin2001} which contains natural images and images of every day scenes and objects. 
    
    The training images were split into patches of size $8\times8$ pixels, resulting in a training set of $2 \cdot 10^{6}$ patches which were used to fit a GMM with 200 components. The training patches were mean subtracted before the GMM was fit. The model is provided by the authors as part of the source code for download\footnote{\url{https://people.csail.mit.edu/danielzoran/epllcode.zip}}. More details of how the model was trained are given in \citet{Zoran2011}.
    
    \subsubsection{GLEAM Radio Data}
    \label{sssec:gleam-radio-data}
    As an alternative we consider a GMM prior learned from astronomical radio data. For this we relied on high signal to noise radio data from the \textit{GLEAM} survey \citep{HurleyWalker2022}. We downloaded the data using the Gleam VO Client\footnote{\url{https://github.com/ICRAR/gleamvo-client}}. We restricted the choice of data to the Galactic Plane between $b=\pm10\deg$ and $\ell = \pm180\deg$ to avoid a bias towards extra-galactic point sources and to represent structures of Galactic sources with rich morphology, such as supernova remnants, pulsar winds etc. The dataset includes Galactic as well as extra-galactic point sources in the background. In total we extracted $\approx 6 \times 10^5$ patches and trained a GMM with 128 components. We used the patch normalization described in Section~\ref{ssec:patch-norm}.

    \subsubsection{JWST Cassiopeia A}
\begin{figure}
        \begin{centering}
            \includegraphics[width=\linewidth]{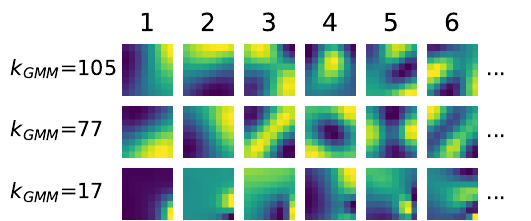}
            \caption{
                Example Eigenimages of the covariance matrices of the GMM learned from the JWST Cas A image. The figure shows the first six Eigenvectors, reshaped as images, of the GMM components $\{105, 77, 17\}$ (same as Figure~\ref{fig:patches}), in descending order of the value of their corresponding Eigenvalue. The color scale varies between the images, so their contrast is different. 
            }
            \label{fig:gmm-eigen-images}
            \script{gmm-eigen-images.py}
        \end{centering}
    \end{figure}

    As a specific prior for the analysis of SNRs we learned a GMM prior from data of the James Webb Space Telescope (JWST). The JWST is a large, infrared space telescope launched in late 2021. It delivers image data at very high spatial resolution with an excellent signal to noise ratio. Both of these properties make JWST data an excellent choice for extracting patches for training a GMM. To limit the data reduction effort we relied on a promotional image of the SNR \textit{Cassiopeia A}\footnote{We downloaded the full resolution (4k)  PNG image from \url{https://webbtelescope.org/contents/media/images/2023/121/01GWQBBY77MHGFV3M3N63KDCEJ}}. We first converted the image to gray scale using a perceived brightness model from the Scikit-Image package \citep{scikit-image}. Then we split the image into $\approx 2$ million $8\times 8$ patches and learned a GMM with 128 mixture components. Again we relied on the patch normalization described in Section~\ref{ssec:patch-norm}. Figure~\ref{fig:gmm-eigen-images} shows the principal Eigenimages of the covariance matrices of three example components of the GMM. The patches show a large variety in morphological shapes, including edges, gradients, bright compact spots and more combinations thereof.

    \subsubsection{Chandra SNRs}
    \label{sssec:chandra-snrs}
    To demonstrate the flexibility in the selection of data to train the GMM, we also chose higher signal-to-noise images from the \chandra catalog of supernova remnants (SNRs). We learned a GMM prior from images of the bright \chandra SNRs \textit{Cas A} and \textit{G292.0+01.8}. In total we extracted $\approx 0.5$ million $8\times 8$ patches from each of the images and learned a GMM with 128 components.

    \subsection{Optimization Strategy}
    \label{sec:opt-strategy}
    Maximising the \aposteriori likelihood defined by Equation~\ref{eq:total} with respect to the model parameters $\mathbf{x}$ is a high dimensional optimization problem. We found the convergence greatly improved when the optimizer operates on $\log{\mathbf{x}}$. This is to be expected, as the brightness in pixels of astronomical can span multiple orders of magnitudes. We used the ADAM algorithm \citep{Kingma2014}, a stochastic gradient descent method, without decay and a fixed learning rate of $\mathrm{lr}=10^{-3}$. To further improve the rate of convergence of the optimization process we applied the following strategy:

    \begin{itemize}
        \item[1.] We start from a random image where each pixel was drawn from a normal distribution with the mean corresponding to the total mean of the flux in the image and a standard deviation of 10\% of the mean. We first estimate $\mathbf{x}$ by using a uniform prior and early stopping after 50 iterations. This corresponds approximately to a reconstruction obtained with the standard RL algorithm. 
        \item[2.] We freeze the image parameters and optimize the parameters associated with the systematic error described in Equation~\ref{eq:model-npred-calibration}.
        \item[3.] Finally we simultaneously optimize
        the image $\mathbf{x}$, the GMM component indicators for each patch, $k_n$ and the parameters associated with systematic error simultaneously until convergence. Convergence is verified visually by inspecting the time series of the log posterior distribution evaluated at the iterates or the models parameters. We adjust the number of epochs (number of passes through the whole set of observations) such that the curve had flattened out for the last few hundred iterations.
    \end{itemize}
    
    \section{Experiments}
    \subsection{Test Datasets}
    \label{subsec:test-datasets}
\begin{figure*}
        \centering\includegraphics{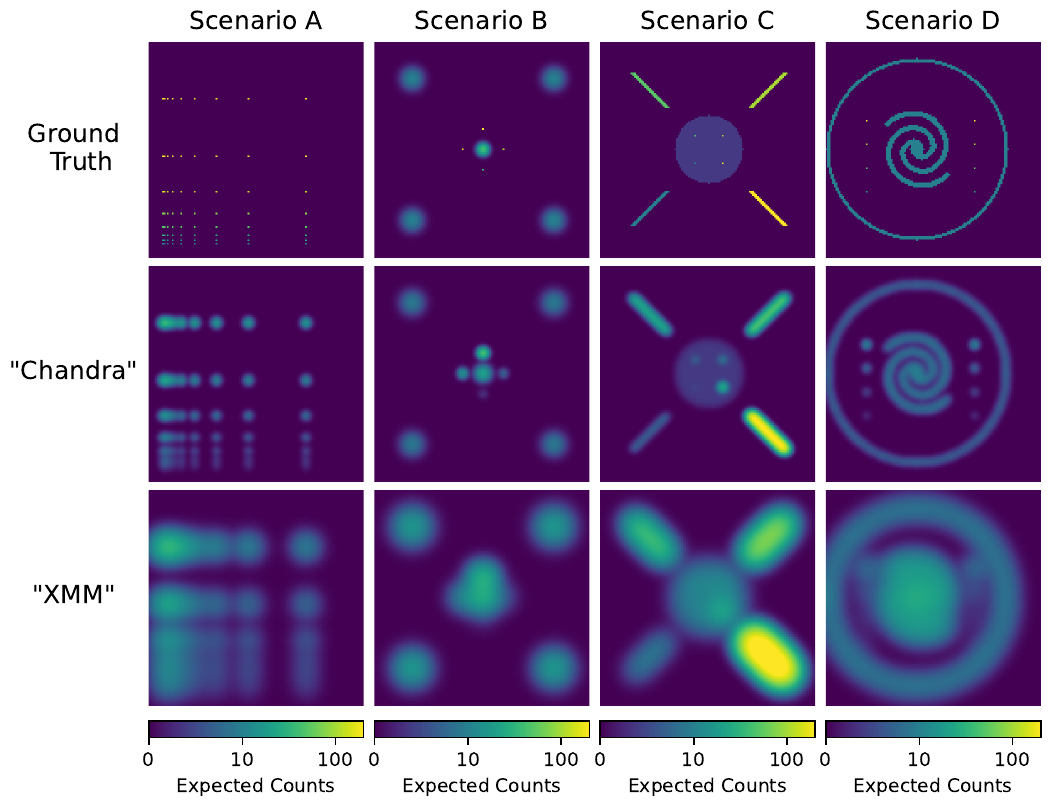}
        \caption{Illustration of the expected counts for the different source scenarios we used to generate test datasets. The source patterns shown are for the {\sl (A)}  {\tt points}, {\sl (B) {\tt asterism}}, {\sl (C)} {\tt shield}, and {\sl (D)} {\tt spiral} cases (see Section~\ref{sec:app:A}. For the \chandra instrument scenario we convolve the ground truth image with a narrow Gaussian shaped PSF of width $\sigma=2$~pix. For the \xmm scenario we used a Gaussian shaped PSF with a width of $\sigma=6$~pix and an effective area that is five times larger than that of \chandranospace. All images are of size $128{\times}128$ pixels.
        }
        \label{fig:scenarios}
        \script{scenarios.py}
    \end{figure*}

    We have designed multiple test scenarios to challenge the several deconvolution algorithms listed in Section~\ref{subsec:methods} and to evaluate and compare their performance. The test scenarios consist of mixtures of point and extended sources, with typical morphological characteristics of astrophysical sources. We arranged the sources in different spatial patterns and with different spacings between them. We include combinations of point and extended sources, sometimes superposed, in four arrangements (labeled A--D in Figure~\ref{fig:scenarios}). Each arrangement is considered at several signal-to-noise ratios (obtained with a constant background and numbered in order of increasing signal). The resulting scenarios are denoted by concatenating the two labels, e.g., A3 or C1, see Appendix~\ref{sec:app:A} for details.  The goal of the experiments is not to set up or devise a \enquote{perfect} deconvolution algorithm, but to help assess their relative strengths and limitations,
    and assess the situations where they fail.
    
    In addition we adopt two distinct instrument scenarios: one which mimics \textit{Chandra} observations with a sharp PSF (a Gaussian with $\sigma=2$~pix) and low background (at 0.01, 0.1, and 1 ct~pix$^{-1}$) but lower effective area (nominally with an exposure of 1 unit), and the other which mimics XMM-\textit{Newton} observations with a broader PSF ($\sigma=3$~pix), higher background (at 0.1, 1, and 10 ct~pix$^{-1}$), and higher effective area (imitated with an exposure of 5 units).
        
    The ground truth and corresponding expected counts for several exemplar combinations of source and instrument scenarios are shown in Figure~\ref{fig:scenarios}. 
    
    \subsection{Methods}
    \label{subsec:methods}
    We compared the performance of \jolideco with different prior assumptions and against the \textit{LIRA} method introduced by \cite{Esch2004}. More specifically we compared the following methods:

    \begin{enumerate}

        \item[\bf M1.] \textbf{Pylira}: we used the LIRA method \citep{Esch2004} via the Pylira Python wrapper \citep{Donath2022}. We relied on the default prior definition and kept the background norm fixed. In total we ran the MCMC sampler for 5000 iterations and discarded the first 500 iterations as a \enquote{burn-in} period to allow the MCMC chain to converge. We computed the mean of the posterior distribution ignoring the burn-in period.
        
        \item[\bf M2.] \textbf{\jolideco (Uniform, n=10)}: we used \jolideco with a uniform prior on $\mathbf{x}$ and ran the optimizer for 10 iterations. We consider this as a baseline method reference similar to RL. While the optimization procedure is different, we expect similar behaviour for the local correlations of pixels. This ad hoc approach of early stopping is often used in practice to regularise the RL output.
        
        \item[\bf M3.] \textbf{\jolideco (Uniform, n=1000)}:  we used \jolideco with a uniform prior on $\mathbf{x}$, but running the optimizer for 1000 iterations. This is the same as above but without early stopping. This gives rise to the issue of RL decomposing extended emission into point sources.
        
        \item[\bf M4.] \textbf{\jolideco (Patch, GLEAM v0.1)}: we used \jolideco with a patch prior learned on the GLEAM data, described in Section~\ref{sssec:gleam-radio-data} and the patch normalization described by Equation~\ref{eq:patch-norm}. We used the optimization strategy described in Section~\ref{sec:opt-strategy} with 2000 epochs.
        
        \item[\bf M5.] \textbf{\jolideco (Patch, Zoran-Weiss)}: we used \jolideco with the patch prior provided by \cite{Zoran2011}. To be consistent with how the GMM was learned we only subtracted the mean of the patch and skipped the division by the mean we introduced in Equation~\ref{eq:patch-norm}. To make the prior applicable to astronomical images, we re-scaled the image to the range $0 > \mathbf{x}_i > 1$ by dividing them by the maximum value defined by the ground truth\footnote{This is not completely fair, as it requires knowledge of the brightest expected pixel in $\mathbf{x}$. In practice this could be estimated from the brightest pixel in the data and the shape of the PSF.}.
    \end{enumerate}

    We ran the comparison on all combinations of methods, source scenarios, background levels, and instrument scenarios. Here we only discuss the subset of the results that we consider the most relevant. The complete set of results is available at \url{https://jolideco.github.io/jolideco-comparison}.

    \subsection{Results}
    \subsubsection{Comparing Methods}
    \begin{figure*}
        \begin{centering}
            \includegraphics[width=\linewidth]{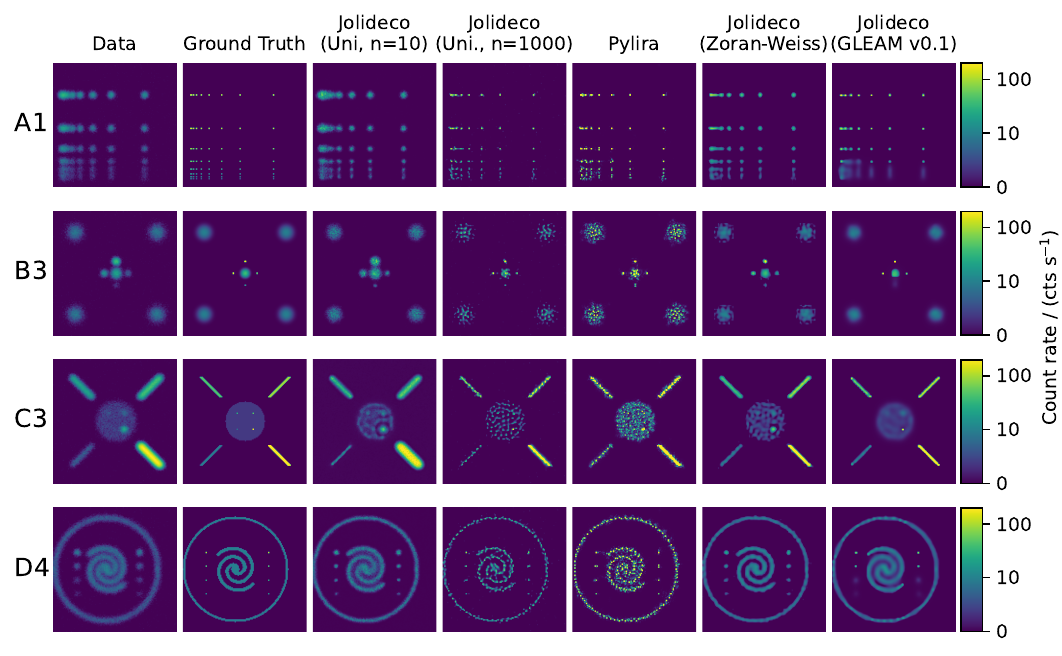}
            \caption{
                Comparison of the different deconvolution methods and prior assumptions for selected source scenarios, as described in Section~\ref{subsec:test-datasets}. The simulations depicted used a fixed background level of $\lambda_{Bkg} = \qty[mode = math]{0.01}{cts/pix}$, a uniform exposure, and a Gaussian PSF of width $\sigma_{PSF} = \qty[mode = math]{2}{pix}$, corresponding to the \enquote{Chandra} instrument scenario. The leftmost column shows the simulated counts and the second column shows the underlying ground truth. The remaining columns show the reconstructed flux for each method and prior assumptions for \jolideconospace. To enhance weak structures the images use an $\arcsinh$ stretching with a scale parameter of $a=0.02$. The stretching is the same for all images. The methods are described in detail in Section~\ref{subsec:methods}. A more detailed representation of the results is available at \url{https://jolideco.github.io/jolideco-comparison}.
            }
            \label{fig:comparison-scenarios}
            \script{comparison-scenarios.py}
        \end{centering}
    \end{figure*}
    First we compare the performance of the five methods and their differing prior assumptions on the different source scenarios. Figure~\ref{fig:comparison-scenarios} shows the results for a selected subset of scenarios A1, B3, C3 and D4. For quantitative comparison we list the corresponding values of the structural similarity index (SSI, \cite{Wang2004}) and the normalized root mean-squared error (NRMSE)\footnote{We use the definition of $NRMSE = \frac{||\mathbf{x}_{GT} - \mathbf{x}||}{||\mathbf{x}_{GT}||}$, where $||$ is the Frobenius norm.} in Table~\ref{tab:metrics}.
    
    For scenario A1 all methods provide acceptable to excellent results for the isolated point sources in the upper right of the image. This demonstrates all methods take into account the PSF and background models correctly. The exception is method M2 (\jolideconospace, uniform prior, stopped early) 
    because its optimization process has not finished. M3, which optimizes for 1000 iterations, clearly improves the result and shows almost perfect point sources corresponding to the size of approximately one pixel. The \jolideco method with the Zoran-Weiss  patch prior (M4) results in less sharp images, because the prior has been learned on natural images of landscapes and every day objects and scenes, which typically do not include point sources. The \jolideco method with the prior learned from GLEAM data (M5), which also includes point sources, shows much improved results relative to M4 but is not as sharp as the results obtained with the uniform prior (M3). In the lower left of the image the point sources are fainter and closer together, challenging all methods. The uniform prior optimized \enquote{to end} (M3) shows good results, separating the sources well. All other methods, however, cannot fully resolve these weak and closely situated point sources.
    
    For scenario B3 all methods recover the point sources well.  There is, however, a large difference  in the reconstruction of the Gaussian shaped extended sources. As expected, the uniform prior assumption lets the emission decompose into many point sources, neglecting the spatial correlations among them. Pylira shows slightly improved results with the gaps between the bright emission peeks being more filled. However, Pylira lacks a convincing result of reconstructing the source into a single object. The \jolideco method with the patch priors shows the best results both visually and as measured by SSI, see Table~\ref{tab:metrics}. With the GLEAM v0.1 patch prior the reconstruction is almost perfect, visually almost indistinguishable from the ground truth. The Zoran \& Weiss prior also shows good reconstruction of the extended sources, but less smooth and leaving a more visible extension in the surrounding point sources.
    \begin{table*}
    \centering
        \begin{tabular}{ c|c|c|c|c|c } 
            Scenario & \thead[c]{Jolideco\\(Uni, n=10)} & \thead[c]{Jolideco\\(Uni., n=1000)} & Pylira & \thead[c]{Jolideco\\(Zoran-Weiss)} & \thead[c]{Jolideco\\(GLEAM v0.1)} \\
            \hline
            A1 & 0.91 / 0.94 & 0.98 / 0.52 & 0.97 / 0.68 & 0.93 / 0.82 & 0.95 / 0.73 \\
            B3 & 0.99 / 0.98 & 0.99 / 0.32 & 0.99 / 0.43 & 0.99 / 0.68 & 1.00 / 0.51 \\
            C3 & 0.95 / 0.76 & 0.96 / 0.83 & 0.95 / 0.80 & 0.98 / 0.36 & 0.98 / 0.56 \\
            D4 & 0.94 / 0.75 & 0.85 / 1.21 & 0.81 / 1.65 & 0.96 / 0.68 & 0.94 / 0.60 \\
            \hline
        \end{tabular}
        \label{tab:metrics}
        \script{metrics-table.py}
        \caption{Image metrics for the different source scenarios and methods shown in Figure~\ref{fig:comparison-scenarios}. The first number in a cell shows the structural similarity index (SSI, the closer to unity, the better) and the second number shows the normalized root mean square (NRMSE, the lower the better).}
    \end{table*}
    The weak extended source with four embedded brighter point sources of Scenario C3 is challenging for all the methods. While the methods again reconstruct the point sources well, the visual differences are large for the weak underlying extended emission. The uniform prior with 1000 iterations decomposes the extended source into a grid of almost equidistant point sources resembling only the outer boundary of the original source. Pylira shows a similar tendency but improves the reconstruction of the extended source by leaving more emission in between the bright emission spots. The best results are achieved by the patch priors, which reconstruct a plausible extended source with approximately uniform brightness. The jet like features are decomposed into a \enquote{trail} of point sources by the uniform prior and Pylira method. The patch prior achieves a much more plausible reconstruction with almost uniform brightness for the linear structures. We also note that the patch prior does not show a preferred direction or bias in reconstruction, that could be possibly encoded in the training data.
    
    The last scenario D4 shows a similar result as the previous test: the uniform prior and Pylira decompose the spiral and ring like structure of the image into points. The patch priors reconstruct a visually more plausible flux image, much closer to the ground truth.

    In summary, \jolideco with the patch priors shows excellent performance on a variety of morphological shapes. We find the prior learned from the GLEAM data to be a very good compromise between the ability to separate nearby point sources and the ability to reconstruct the spatial correlation of extended sources.
    
    \subsubsection{Comparing Signal-to-Noise Scenarios}
    \begin{figure*}
        \begin{centering}
            \includegraphics[width=\linewidth]{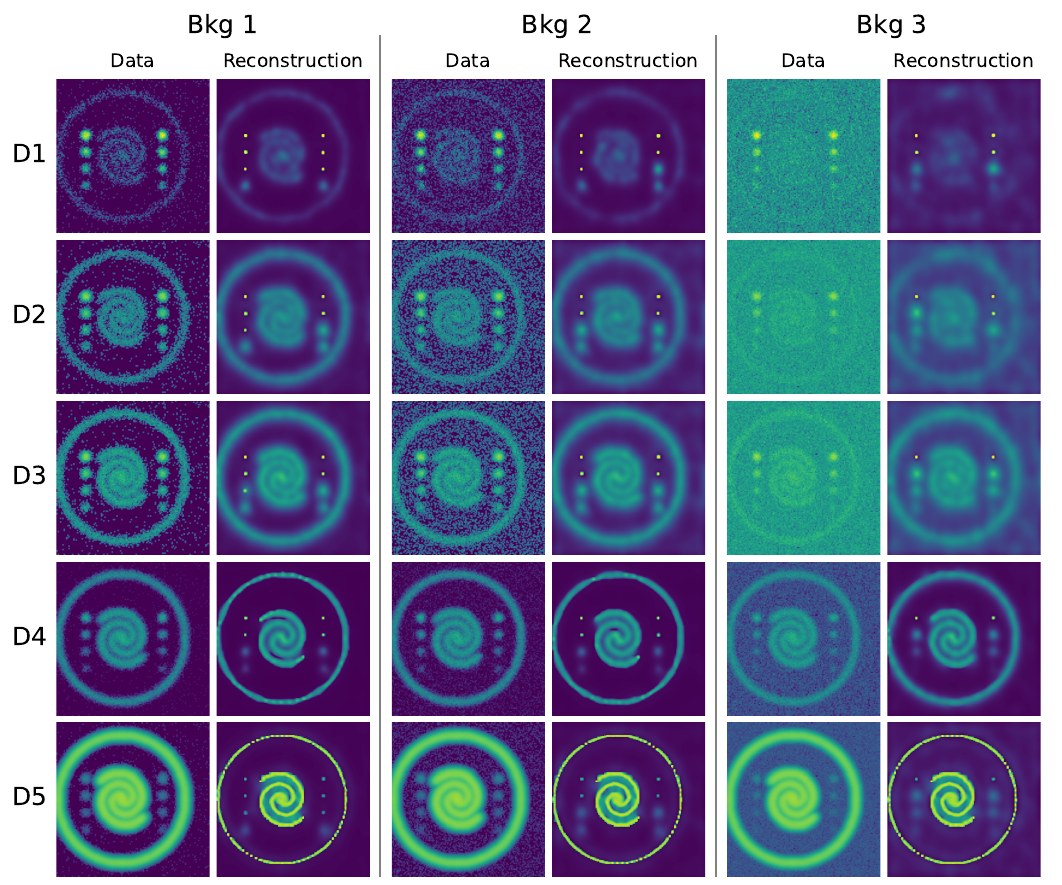}
            \caption{
                Comparison of \jolideco with the GLEAM v0.1 prior for different background levels and signal strength, as represented by scenario D1-D5. The images are grouped by background level with each pair of columns showing the counts data in the left column and the reconstruction in the right column. The rows show the scenarios D1-D5 as described in Section~\ref{subsec:test-datasets}. Each pair of columns corresponds to a  background level,  \qty[mode = text]{0.001}{cts / pixel}, \qty[mode = text]{0.01}{cts / pixel} and \qty[mode = text]{0.1}{cts / pixel}. The complete results for the simulations and additional information are available at \url{https://jolideco.github.io/jolideco-comparison}.
            }
            \label{fig:comparison-signal-noise}
            \script{comparison-signal-noise.py}
        \end{centering}
    \end{figure*}
    As a second experiment we explore the dependency of the results on the background level and signal strength. Figure~\ref{fig:comparison-signal-noise} shows the results for the \jolideco method with the GLEAM prior in Scenario D1-D5. The background level varies between $\lambda_{Bkg}= 0.01, 0.1$ and \qty[mode = text]{1}{counts/pixel}. As expected, the reconstruction quality improves with higher signal strength and lower background level. Visually the method converges to the ground truth for very high signal-to-noise ratio. Indeed with very high signal strength there is no dependency of the result on the background level, particularly in Scenario~D5. On the other hand, the method struggles to reconstruct a useful signal in the case of very high background (scenarios D1-D3).

    \subsubsection{Comparing Instrument Scenarios}
    \begin{figure*}
        \begin{centering}
            \includegraphics[width=\linewidth]{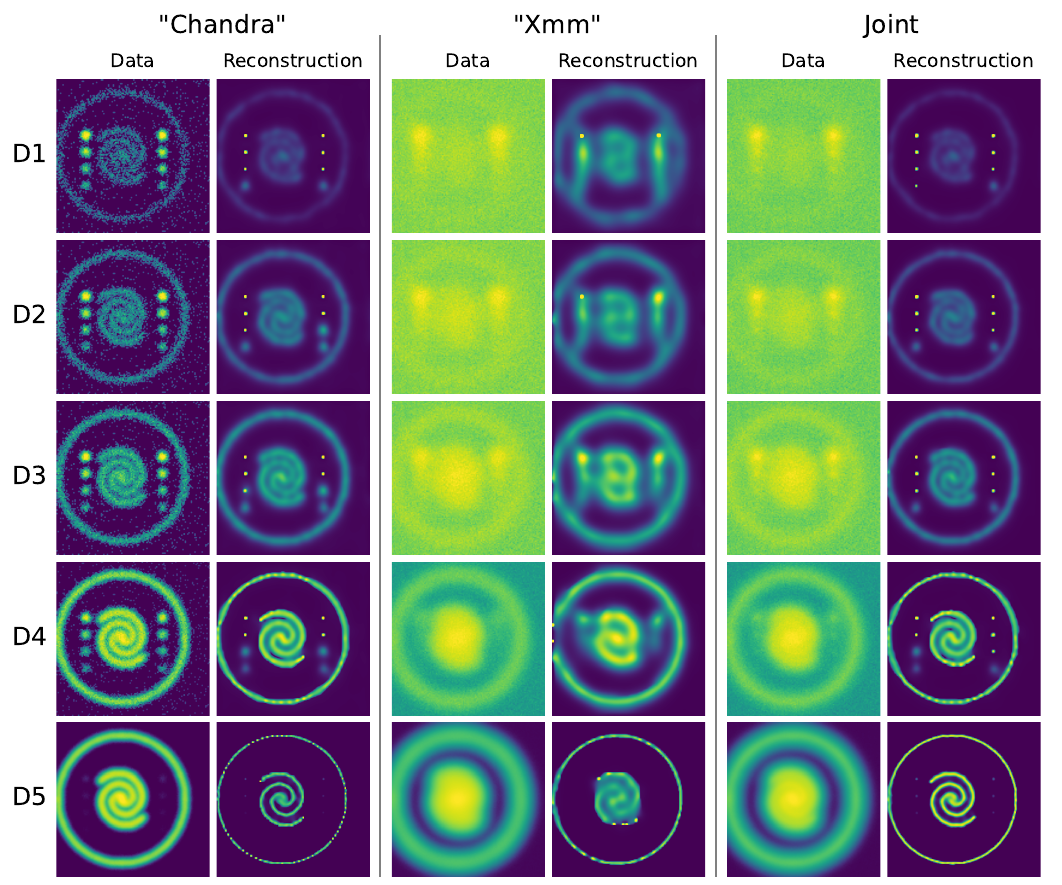}
            \caption{
                Comparison of \jolideco with the GLEAM v0.1 prior for the different instrument scenarios. The images are grouped by instrument scenario with each pair of columns showing the counts data in the left column and the reconstruction in the right column. The rows show the scenarios D1-D5 as described in Section\ref{subsec:test-datasets}. 
                All results are available at \url{https://jolideco.github.io/jolideco-comparison}.
            }
            \label{fig:comparison-instruments}
            \script{comparison-instruments.py}
        \end{centering}
    \end{figure*}
    Finally we compare \jolideco results  for the two instrument scenarios. Figure~\ref{fig:comparison-instruments} shows the data as well as flux images reconstructed using \jolideco and the GLEAM v0.1 patch prior. We consider source Scenarios~D1-D5 and the  \enquote{Chandra} and \enquote{XMM} instrument scenarios as well as using a combination of their data to reconstruct a single image. 

    Using the data from the \enquote{Chandra} scenario alone results in a good reconstruction in all variations of scenario D. As expected the reconstruction quality improves for higher signal to noise ratio. 
    
    Using only the data from the \enquote{XMM} scenario results in lower quality reconstructions, due to the lower angular resolution of the data. However, we find that the spiral arm structure is reconstructed at all levels of signal strength, even when barely visible by eye. The ring structure is sharpened considerably in all cases as well. The near-by point sources on the left and right of the spiral arm structure cannot be separated into individual point sources, but remain blurred along the direction of their neighbours. This a similar situation as in the lower left of the image in the A1 point source scenario, when sources are to close by.
    
    Combining the data results in the best reconstruction quality. The addition of the lower resolution \enquote{XMM} data does not negatively affect the quality of the overall reconstruction. When comparing Scenario~D5 we see that the \enquote{Chandra} reconstruction shows a very good reconstruction of the spiral structure but shows speckles in the surrounding ring. The latter suggests that the GMM prior is not strong enough to maintain the pixel to pixel correlations in the case of high signal strength. The strength of the prior could be increased with the $\beta$ parameter, but for this study we use the default settings and do not tweak individual parameters. The speckly behaviour is not observed in the \enquote{XMM} case, where the reconstruction of the ring is smoother. In this case the correlation between pixels that are further apart is stronger because of the larger correlation PSF. However, the reconstruction of the spiral structure in the center shows visual artifacts, with the individual arms not completely separated. The joint reconstruction shows the positive qualities of both individual reconstructions. This confirms the results from \cite{Ingaramo2014}, where the authors found that combining multiple measurements for deconvolution with RL to be a \textquote{general tool for combining images with complementary strengths}.

    \section{Application Examples}
    \begin{figure*}
        \begin{centering}
            \includegraphics[width=\linewidth]{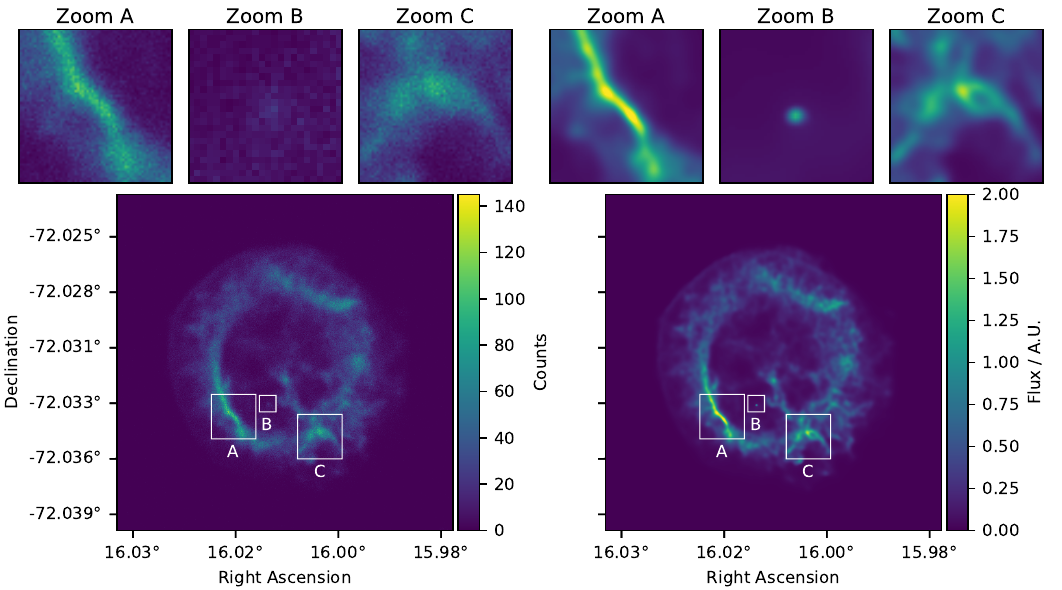}
            \caption{
               \jolideco result with the JWST patch prior for \chandra observations of SNR \textit{1E~0102.2–7219}. The bottom left image shows the summed counts from 25 observations. The bottom right image  shows the the \jolideco deconvolution. The flux is given as a counts rate relative to obsID 1308. The panels \textit{Zoom A}, \textit{Zoom B} and \textit{Zoom C} show a zoomed-in version of three example regions in the SNR, marked with the white rectangles in the large images. The colormap and scale used in the zoom insets is the same as used for the larger images.
            }
            \label{fig:example-chandra}
            \script{example-chandra.py}
        \end{centering}
    \end{figure*}
    \subsection{Deep \chandra Observation}
    \label{sec:chandra-example}
    \subsubsection{Image Example}
    We apply \jolideco to a set of short-exposure \chandra observations of the SNR \textit{1E~0102.2–7219} and reconstruct a single long-exposure image using the patch prior learned from JWST data. In total, we chose 25 observations taken with  the ACIS-S instrument and no grating. We chose the data taken after 2006 when the contamination layer results in lower pileup. The list of selected observations including their exposure duration is shown in Table~\ref{tab:chandra_obs} of Appendix~\ref{sec:chandra-table}.

    We reduced the data using CIAO v4.15 \citep{Fruscione2006} and a custom Snakemake workflow (see also Section~\ref{sec:reproducibility} on reproducibility). For the counts image we selected an energy range from \qty[mode = text]{0.5}{keV} to  \qty[mode = text]{7}{keV} and a pixel size of $0.25$ of the native pixel size of \chandranospace ACIS detector. This corresponds to an absolute  pixel size of $\approx$~\qty[mode = text]{0.125}{$\arcsec$/pixel} or a sampling of $\approx2$ pixels for the $38\%$ containment radius of \chandranospace's PSF. \chandranospace's instrument response varies with the observation offset and energy \citep{ChandraPOG2022}, so we simulated the PSF for each observation independently using the \texttt{marx} tool \citep{Davis2012} and the \texttt{simulate\_psf} script. As a reference spectrum for the PSF simulation we assumed the model derived by \cite{Plucinsky2017}. We converted the spectral script from XSPEC to {\it Sherpa} using the helper script \texttt{convert\_xspec\_script}.

    We assume a spatially constant background of \qty[mode = text]{1e-5}{cts/pixel} but allow the normalization to vary in the reconstruction process. We assume the exposure to be spatially constant and express its absolute value for each observation as the ratio of observation time with respect to a reference observation, namely obsID 8365. For \jolideco we use the GMM prior learned on the JWST Cas A image. As there are many counts in the image we additionally use an oversampling factor of 2, meaning the image $\mathbf{x}$ is computed on a finer grid and downsampled after convolution with the PSF to the same grid as the counts data. The patch prior is also evaluated on the finer grid.
    
    We allow the background level and position calibration (cf., Equation~\ref{eq:model-npred-calibration}) to vary among the observations. As a reference for the positional shift we also use obsID 8365. Figure~\ref{fig:example-chandra} shows the final result. \jolideco clearly both improves the angular resolution and reduces the noise of the counts image. 
    
    \textbf{Zoom A} of Figure~\ref{fig:example-chandra} details an example of a filament structure. \jolideco maintains the extended structure along the filament, while sharpening in the perpendicular direction. Looking at the upper right corner one can also see that weak background emission is suppressed, improving the contrast of the image.
    
    \textbf{Zoom B} shows the region around the putative central compact object of the SNR (\cite{Vogt2018}, \cite{Hebbar2020}, but \cite{Long2020}). While the emission is barely visible in the counts image it is clearly evident above the background in the \jolideco reconstruction. The reconstruction leaves a visible extension in the point-like source. However, at this level of analysis and signal to noise, both scenarios remain plausible: a real intrinsic extension of the feature, as well as a remaining reconstructed extension in a real point source because of low signal to noise\footnote{We note that, as illustrated in Section~\ref{subsec:methods} for Scenario~A, the remaining extension of the point source after deconvolution may be smaller if the source is brighter.} 
    
    To draw stronger scientific conclusions a dedicated analysis of the region, including a simulation of the false detection ratio of point sources would be needed. We estimate the position of the feature by finding the brightest pixel in the map, obtaining $\alpha_{J2000}=01^h04^m02.75^s$ and $\delta_{J2000}=-72^d02^m00.19^s$ which is consistent with the value reported by \cite{Long2020}. This demonstrates an absolute feature position that was reconstructed with \jolideco that is consistent with the existing literature.
    
    \textbf{Zoom C} shows a circular filament structure of the SNR. In the counts data the structure is barely visible, while the \jolideco reconstruction shows a distinct circular structure with a locus of bright emission. This demonstrates the ability of \jolideco to reconstruct sub PSF scale structures with astrophysical relevance.

    Checking the residual images per observation (not shown here), we notice that the assumption of time independence of the source and the instrument response is not completely fulfilled in this case. The detection efficiency of \chandra changed and the sub-structure in the remnant expanded over the time of the observations. Both effects are previously known and have been reported for example in \cite{Xi2019}. Despite the clear improvement in spatial resolution, systematic issues such as the proper-motion of structures and calibration uncertainties, might limit the quality of the result. The full analysis example is available at \url{https://github.com/jolideco/jolideco-chandra-examples}.

    \begin{figure*}
        \begin{centering}
            \includegraphics[width=\linewidth]{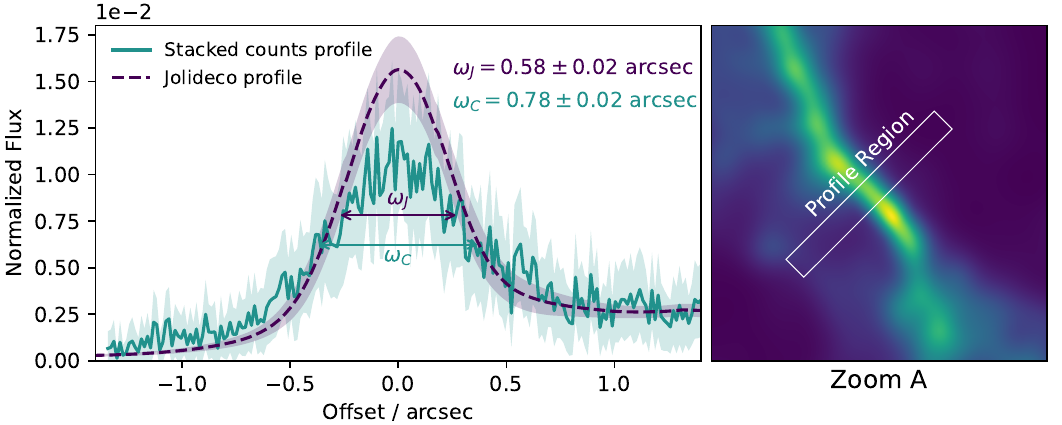}
            \caption{
                The left panel shows the deconvolved flux and counts profile of the region illustrated in the right panel. The thick solid and dashed lines represent the mean across the 100 bootstrap-resampled datasets. The counts profile for each bootstrapped dataset is summed (stacked) across the 25 observations per dataset, while the \jolideco profile is computed from a single joint reconstruction based on all 25 observations. The variability across the 100 bootstrapped datasets is represented by the $3\sigma$ error bands, shown transparent in the background for both profiles. Both profiles are normalized such that they integrate to unity and are aligned such that the position of their peak coincides. 
            }
            \label{fig:chandra-e0102-zoom-a}
            \script{chandra-e0102-zoom-a.py}
        \end{centering}
    \end{figure*}
    
    \subsubsection{Flux Profile}
    To quantify the improvement in angular resolution of the reconstructed image over the raw data we compute a flux profile on the \jolideco (with the JWST patch prior) reconstruction of E0102. We chose a region perpendicular to the filament shown in Zoom~A of Figure~\ref{fig:example-chandra} of the size $128\times128$ pixels. We compute errors via the bootstrap, with resampled data obtained by sampling the event lists with replacement for each individual observation. For each resampled dataset (composed of 25 observations) we run \jolideco and compute a binned flux profile on the reconstructed image. In total we sample 100 different realizations, where each reconstruction runs for about 10 minutes on a standard M1 CPU. In this way we obtain a set of profiles, which we use to compute the mean and error per bin of the profile. For comparison we also compute the profiles and associated errors directly on the bootstrapped counts summed across the 25 observations (without processing or fitting). The result is shown in Figure~\ref{fig:chandra-e0102-zoom-a}.

    For each bootstrapped dataset, we then compute the full-width at half-max (FWHM) of the profiles by fitting a Gaussian shape with a linear gradient as a baseline, for both the counts profile ($\omega_C=0.78{\pm}0.02$~arcsec) and the \jolideconospace-deconvolved intensities ($\omega_J=0.58{\pm}0.02$~\arcsec), where the uncertainty is represented by the standard deviations in the bootstrap sample.  Note that a significant narrowing of the cross-sectional profile is seen.  The width of the filament however is definitively $>0''$, suggesting that it is resolved in the \jolideconospace\ reconstruction.\footnote{The projected radial profiles of the on-axis \chandra\ PSF (see the \chandra/MARX raytrace assessment at \url{https://space.mit.edu/cxc/marx/tests/PSF.html}) show that the 39\% enclosed counts fraction radius is $\approx$0.26~arcsec.  Approximating the PSF as a 2D Gaussian, this corresponds to a $1\sigma$ radius, and a FWHM $\omega_{\textrm{PSF}}{\approx}0.62$~arcsec.  Since the observed profile is expected to result from a convolution of the PSF with the true profile, we expect
    $$\omega_C^2 = \omega_{\textrm{PSF}}^2+\omega_J^2 \,.$$  
    A true filament width of $\gtrsim$0.5~arcsec is thus necessary to match the measured $\omega_C$.  That the estimated $\omega_J$ is larger by $\approx$0.1~arcsec can be attributed to presence of several systematic confounding factors, such as the PSF being heavier tailed than a Gaussian, and the intrinsic profile of the filament not being Gaussian.}
    
    \subsection{\fermi Event Types}
    \label{sec:fermi-lat-example}
     As a second application example, we apply \jolideco  to data from the Fermi Large Area Telescope (LAT) \citep{Atwood2009}. The LAT is a pair conversion \gammaray telescope on board of the Fermi satellite. It operates in the energy range between \qty[mode = text]{30}{MeV} to  \qty[mode = text]{300}{GeV}. For this example we chose  a Galactic source with complex morphology: the supernova remnant Vela Junior. We use 14 years (from 2008 August 4 to 2023 August 4) of \enquote{P8R3\_SOURCE} photons with reconstructed energy in the range of \qty[mode = text]{10}{GeV} to \qty[mode = text]{2}{TeV} and applied a zenith angle cut of \ang{105}. This is consistent with the data selection range in \cite{Ackermann2017}. To preserve as much information as possible on the angular resolution of the instrument, we split the data into the four event types \enquote{PSF0}, \enquote{PSF1}, \enquote{PSF2} and \enquote{PSF3}, as defined by the \fermi \enquote{Pass8} data release\footnote{\url{https://fermi.gsfc.nasa.gov/ssc/data/analysis/documentation/Pass8_usage.html}}. Where PSF0 represents the subset of data with the worst and PSF3 the best angular resolution. Instead of different observations, the joint likelihood is computed across those different event types.
     
     We chose a pixel size of \qty[mode = text]{0.02}{$\deg$}, which corresponds to approximately 1/5 of the size of the PSF in the best event class. For the first part of the data reduction we use the standard \enquote{Fermitools} v2.2.0 \citep{Fermitools2019}.
    
    The Fermitools produce energy dependent counts, exposure and PSF cubes. To further reduce the data to images we use the Gammapy package v1.0 \citep{GammapyZenodov1.0.1, Gammapy2023}. To compute an image of the expected background emission we rely on the latest Galactic diffuse emission model \texttt{gll\_iem\_v07.fits} and also include nearby sources from the 3FHL catalog \citep{Ajello2017}. We first fit the norm and spectral indices of all the 3FHL sources and the background model to the data using Gammapy, excluding a circular region of radius \qty[mode = text]{1.02}{$\deg$} centered on Vela Junior. Finally we reduce this model to an image by summing over the energy dimension. To compute the exposure and PSF image we assume a power law with index $\Gamma=1.77$ \citep{Ajello2017} as a spectral model for Vela Junior and integrate over the energy dimension of the cube weighting according to this spectral model.
    
\begin{figure*}
        \begin{centering}
            \includegraphics[width=\linewidth]{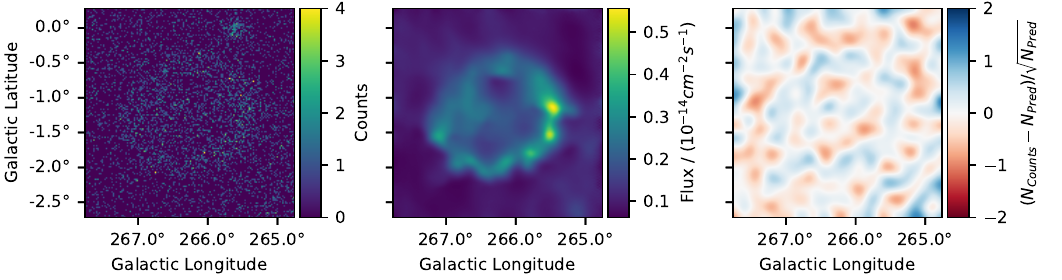}
            \caption{
                Illustration of the \jolideco result for the \fermi data of the supernova remnant \textit{RX J0852.0-4622} or \textit{Vela Junior}. The left image shows the counts above \qty[mode = text]{10}{GeV}. The different event classes are stacked into a single image with bin size $0.02$~degrees. The image in the center shows the flux reconstructed by the \jolideco method. The image on the right shows the standardized residuals as computed by Equation~\ref{eq:approx-sigma} and smoothed with a \textit{Gaussian} kernel of width 5~pixels $\approx 0.1$~degrees, similar to the size of the PSF.
                More information on this analysis example can be found on \url{https://github.com/jolideco/jolideco-fermi-examples}.
            }
            \label{fig:example-fermi-lat}
            \script{example-fermi-lat.py}
        \end{centering}
    \end{figure*}

    We chose the GMM patch prior learned from \chandra SNRs described in the Section~\ref{sssec:chandra-snrs}. Because of the limited statistics we do not use oversampling but keep the pixel size of the counts image. Figure~\ref{fig:example-fermi-lat} shows the data,  the \jolideco results, and corresponding residuals. The \fermi data is sparse, with limited signal counts. \jolideco clearly improves the noise level as well as angular resolution compared to the data. The calibration parameters of the predicted counts converged to different norms for the background model and positional shifts of order of $<1$ pixel for the different event types. In this way \jolideco corrects for systematic errors in the reconstruction of the position. The reconstruction includes background emission around Vela Junior, indicating that the Galactic diffuse background model is either underestimated in brightness and/or associated with systematic errors. The background estimate could be improved by choosing a larger region of interest. However this would also require a larger effort of modeling nearby sources, which we avoided here.
    
    To further check the quality of the reconstructed flux we compute a map of standardized residuals defined by:

    \begin{equation}
        \label{eq:approx-sigma}
        \rho(N_{\mathrm{Data}}, N_{\mathrm{Pred}}) \approx \frac{1}{\sqrt{\pi \sigma^2}}\frac{(N_{\mathrm{Data}} - N_{\mathrm{Pred}}) \circledast g_{\sigma}} {\sqrt{N_{\mathrm{Pred}}\circledast g_{\sigma}}}.
    \end{equation}
    
    where $g_{\sigma}$ is a Gaussian smoothing kernel of width $\sigma=5$ pixels. The result is shown in the right panel of Figure~\ref{fig:example-fermi-lat}. The image shows no systematic effects and a spatially homogeneous distribution of the residual noise. The resulting distribution of residual significances per pixel is centered at $\mu=-0.001$ with a width of $\sigma=0.991$, indicating a close agreement with a standard normal distribution, as expected for a good model of the emission. In Figure~\ref{fig:example-fermi-lat-non-calibrated} we also show the residuals per PSF class with and without the correction defined by Equation~\ref{eq:model-npred-calibration}. More details of the \fermi analysis are available at \url{https://github.com/jolideco/jolideco-fermi-examples}
    
    \section{Paper Reproducibility}
    \label{sec:reproducibility}

    In order to facilitate reproducibility, we provide the source code for the numerical results
    and figures presented in this article at \url{https://github.com/jolideco/jolideco-paper}. The results in this paper can be reproduced using using the tool \texttt{Showyourwork} \citep{Luger2021} and following the instructions in the \texttt{README.md} file in the same repository. All data are available for download from Zenodo or can be reproduced using the custom  \texttt{Snakemake} workflows and repositories provided on the \jolideco GitHub page\footnote{\url{https://github.com/jolideco}}. For the Fermi-LAT data reduction we used a custom \texttt{Snakemake} workflow, also  available on GitHub\footnote{\url{https://github.com/adonath/snakemake-workflow-fermi-lat}}. For the Chandra data reduction we used a custom \texttt{Snakemake} workflow that is also available on GitHub\footnote{\url{https://github.com/adonath/snakemake-workflow-chandra}}.

    \section{Summary \& Outlook}
    In this article we presented a new method for deconvolution and denoising of images contaminated by Poisson noise. Using multiple observations of the same sky region the method reconstructs a single flux image using the MAP estimate of the joint posterior distribution under a patch-based image prior. We show that the method leads to improved results compared to existing methods in a variety of morphological, signal to noise, and instrument scenarios. Improvement is especially pronounced for extended sources of a size similar to the PSF. In this case, \jolideco can reliably recover features that are well beyond the angular resolution of the instrument. We also show that the approach of jointly analyzing multiple observations leads to improved reconstruction results. Including data with lower angular resolution in the \jolideco reconstruction improves the fitted correlation structure among pixels across larger distances, while not negatively affecting the overall sharpness of the reconstruction.

    We imagine that \jolideco is the first step of a larger project building ML-inspired methods for the reconstruction of astronomical images that are dominated by Poisson noise. While \jolideco has already achieved results that are superior to any other existing method, it still has significant potential for future extension and improvements.
    
    The first category of future improvements is related to the prior handling. For example, more general mixture models such as the Student-t or a generalised Gaussian mixture model can be used to model the distribution of patches. \citet{VanDenOord2014} shows that these more flexible models can account for the tails of the patch distribution and lead to improved results on natural images. We expect these improvements to transfer to astronomical images.

    To allow the patch-based prior distribution to capture correlations at larger scales,
    \cite{Papyan2015} proposed to combine patches extracted at multiple scales of the image when training the GMM. This is a promising approach to further improve the reconstruction quality for extended sources with low signal to noise ratios.
    
    Working in the context of computer tomographic image reconstruction, \cite{Altekrueger2022} used normalising flows (NF) in their formulation of the patch prior in order to further improve the general expressiveness of the prior. As with a GMM, we can evaluate the NF-based prior distribution and thus construct an explicit posterior distribution for optimization and computation of the MAP estimate. \citet{Ulyanov2017} introduced another promising prior distribution, known as the \enquote{Deep Image Prior} and showed that prior structure for an image can be built into a deep neural network as an \enquote{inductive bias}, where the architecture of the network encodes the prior information on the correlation structure among neighbouring pixels. The authors showed, that the deep image prior successfully imposes self similarity on structures in the image. As structures in astronomical images often show sell-similarity this is an interesting approach, that in principle could be combined with the patch prior.

    A second category of improvement to \jolideco is generalizing it to jointly model the energy and spatial domains, thus enabling hyperspectral image reconstruction. Instruments such as \fermi combine instrument responses that vary strongly with energy along with a limited overall energy resolution. This configuration introduces correlations between images in neighbouring energy bands. Accounting for energy data in the reconstruction would allow us to handle these correlations and also open up the possibility of merging multiple \xray datasets in the joint spectral/spatial domain,  thus improving both  angular and energy resolution. 

    \jolideco can already handle an independent background component. However we plan to extend \jolideco to enable it to handle multiple source components at the same time. This allows us in principle, as shown by \cite{Selig2015} and \citet{Pumpe2018} to separate extended sources from point-like sources, based on different prior assumptions on the spatial correlation and distribution of pixels for each model component. 

    A last category of improvement to \jolideco is computing principled error estimates for the reconstruction. In this paper, we use bootstrap methods. Since we can compute the full posterior distribution, however, we could build an approximation to the covariance matrix of the MAP estimate, for example, by using a low-rank or sparse representation.
    Alternatively we can use MCMC to obtain a sampling from the full posterior distribution, enabling us to explore the distribution and quantify  uncertainty both in the parameters and functions of the parameters constructed to represent structures in the image. 
    
    Finally we invite the community of X-ray and \gammaray astronomers to use  \jolideco to combine datasets and make new scientific discoveries, beyond what is possible with the angular resolution of existing instruments. We also welcome contributions and feedback to the public \jolideco code and validation repositories. 

    \section*{Acknowledgements}
    This work was conducted under the auspices of the CHASC International Astrostatistics Center.
    CHASC is supported by NSF grants DMS-21-13615, DMS-21-13397, and DMS-21-13605; by the UK Engineering
    and Physical Sciences Research Council [EP/W015080/1]; and by NASA APRA Grant 80-NSSC21-K0285.
    
    We thank CHASC members for many helpful discussions, especially Xiao-Li Meng and Katy McKeough.
    DvD was also supported in part by a Marie-Skodowska-Curie RISE Grant (H2020-MSCA-RISE-2019-873089)
    provided by the European Commission.
    
    Aneta Siemiginowska, Vinay Kashyap, and Douglas Burke further acknowledge support from NASA
    contract to the Chandra X-ray Center NAS8-03060.

    Some of the computations in this paper were conducted on the Smithsonian High Performance
    Cluster (SI/HPC), Smithsonian Institution \url{https://doi.org/10.25572/SIHPC}.

    This research has made use of data obtained from the Chandra Data Archive and software provided by the Chandra X-ray Center (CXC) in the application packages CIAO \citep{Fruscione2006} and {\it Sherpa} \citep{Freeman2001,Burke2022}.
    
    \newpage
    \bibliography{bib.bib}

    \appendix
    \section{Details on Source Patterns Used in Numerical Experiments}
    
    \label{sec:app:A}

    We have devised a series of elaborate simulation scenarios  consisting of point sources and extended sources of different shapes, sizes, orientations, and spacings in order to test deconvolution algorithms for astronomical low-count images.  There are four distinct patterns, illustrated in Figure~\ref{fig:scenarios}, which we denote (1) {\sl Pointilist}, (2) {\sl Asterism}, (3) {\sl Shield}, and (4) {\sl Spiral}.  The construction of each pattern is described below.  All patterns are placed in a $128{\times}128$ pixel image.

    \subsection{Scenario A1: ``Pointilist"}
    The first scenario conisists of a series of point sources arranged in a grid with increasing horizontal and vertical separations.  The sources placed along a horizontal row all have the same strength, which increases with $y$-axis locations.  The source strengths are 
    $$s_{\rm A} = \{10, 20, 30, 50, 80, 130, 210, 340\} \cdot \tau ~\textrm{counts} $$
    for
    $$y = [8, 10, 13, 18, 26, 39, 55, 76]~\textrm{pix} \,,$$
    where $\tau$ denotes an arbitrary scaling factor, set to 1 for \textit{Chandra}-like simulations and to 5 for XMM-like simulations.
    The $x$-axis locations of the sources for a given $y$ are
    $$x=[8, 9, 10, 11, 13, 16, 21, 29, 42, 61, 95]~\textrm{pix} \,.$$
    The spatial separation and intensity changes are based on the Fibonacci series, which places them on a cadence that is sub-logarithmic but supra-linear, and thus provides a good trade-off between dynamic range and point density. Notice that at the lower left corner of the image, adjacent sources are placed closer together, either in the adjoining pixel or separated by one pixel, with low source strengths, posing an almost impossible challenge to deconvolution algorithms.  In contrast, the upper right corner of the image is sparse, with strong well separated sources.

    \subsection{Scenario B1-B3: ``Asterism"}
    The pattern for these scenarios includes a diffuse Gaussian source ($\sigma=2$~pix) at the center of the image with pixel location (64,64). Four larger extended sources (Gaussians with $\sigma=4$~pix) with the same number of counts as the central source are set at each of the corners.  Four point sources with intensities decreasing by a factor $1/3$ are arranged around the central diffuse source to the top, left, right, and bottom.  The three signal-to-noise ratios (indicated by the scenario number) are determined by the counts in the diffuse sources, all of which are set to one of the values in $$\{B1, B2, B3\} := \{100, 500, 1000\} \cdot \tau ~\textrm{counts}/\textrm{pix}\,,$$ where $\tau=1$ for \chandra-like and $\tau=5$ for \xmm-like simulations, as above.  The diffuse sources are located at
    $$(x,y)_{{\rm diffuse}} = \{(22,22), (22,106), (106,106), (106,22)\} ~\textrm{pix} \,,$$
    and the point sources at
    $$(x,y)_{{\rm point}} = \{(64,76), (44,64), (74,64), (64,44)\} ~\textrm{pix}\,,$$
    with counts
    $$\left\{1, \frac{1}{3}, \frac{1}{3^2}, \frac{1}{3^3} \right\} \times$$
    those in the diffuse sources respectively.  The pattern of point sources explores the effects of contamination by large extended sources, while the diffuse sources at the corners help to characterize the edge effects in deconvolutions.

    \subsection{Scenario C1-C3: ``Shield"}
    These scenarios consists of a flat disk at the image center, with superposed point sources, and a jet-like linear extensions radiating to the corners.  The disk has a radius of 20~pix. The overall signal-to-noise ratio of each of the three scenarios is determined by the disc's brightness, which is set to one of the values in 
    $$\{C1, C2, C3\} := \{0.5, 1, 2\} \cdot \tau ~\textrm{counts~pix}^{-1} \,,$$
    where again $\tau$ represents an arbitrary scaling factor to mimic telescope sensitivity.  Four point sources are placed at positions displaced at $\{\pm8,\pm8\}$~pix from image center, with intensities set to
    $$\{10, 50, 100, 500\} \cdot \tau ~~\textrm{counts}$$
    clockwise starting from the lower left, respectively.  Linear sources, each with a horizontal width of 3~pix, are placed along the $\pm45^o$ lines extending from radial distances of 36~pix to 64~pix from the image center, with surface brightnesses of 
    $$\{10, 50, 100, 500\} \cdot \tau ~\textrm{counts~pix}^{-1} \,,$$
    matching those of the nearest point sources.  This pattern simultaneously tests the ability of deconvolution algorithms to distinguish the disk from the background, and the embedded point sources from the disk.  The linear radiating sources present a different difficult challenge to algorithms which use symmetrical kernels, which usually break such features into a string of point-like sources. (We emphasize that such features are often seen in real astronomical images as jets or detector features like streaks or bad CCD columns.) 

    \subsection{Scenario D1-D5: ``Spiral"}
    These scenarios have 4 major components. First is a double spiral structure, centered at $(-10, 0)$~pix, covering 500$^o$~deg, and with a maximum radial extent of 24~pixels and a width of 3~pixels. Second is a disk of radius 4, centered at $(-10, 0)$~pix.  Third are two vertical rows of point sources located at x-axis positions of 40~pixels and 88~pix, with y-axis locations spanning $\{\pm7,\pm21\}$~pix. Fourth is an annulus that encloses all these elements; it is also centered at $(-10, 0)$~pix, with inner radius 52~pix and outer radius 54~pix.  
    The point sources have intensities of $\{25, 50, 100, 200\}$~counts increasing with vertical height. The overall signal-to-noise ratio of the scenarios is determined by the surface brightnesses of the spiral, disk, and annulus, which are set to one of the values in 
    $$\{D1, D2, D3, D4, D5 \} := \{0.5, 1, 2, 10, 1000\} \cdot \tau ~\textrm{counts}.$$  
    This configuration tests the recovery of closely located diffuse structures (close spiral arms), field-of-view edge effects (annulus that approaches close to the left edge of the image), and the visibility of point sources close to diffuse structures.  The case with the surface brightness of 1000~counts~pix$^{-1}$ tests the asymptotic limit of the high counts case.  Since only the diffuse emission components have the enhanced brightness, this scenario also demonstrates the benefits of a sharp PSF; broad PSFs contaminate the diffuse emission around the point sources which results in a significantly higher difficulty in recovering faint sources.
    
    \section{Additional Tables and Figures}    
\begin{figure*}
        \begin{centering}
            \includegraphics[width=\linewidth]{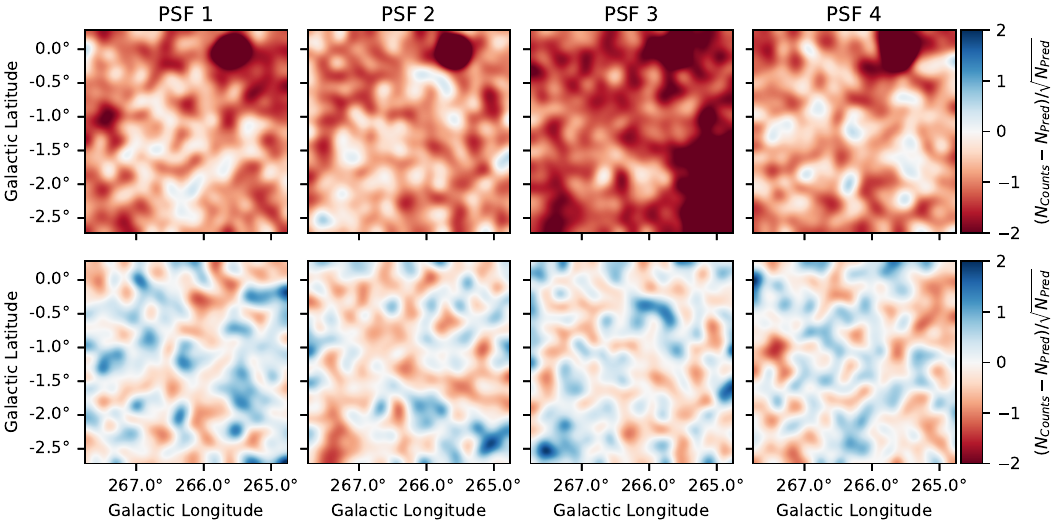}
            \caption{
                Illustration of the residuals of the \fermi analysis example of \textit{RX J0852.0-4622} from Section~\ref{sec:fermi-lat-example}. All images show the standardized residuals as computed by Equation~\ref{eq:approx-sigma} and smoothed with a \textit{Gaussian} kernel of width 5~pixels $\approx 0.1$~degrees, similar to the size of the PSF. The upper row shows the result without calibration, the lower row shows the result with the calibration applied, according to Equation~\ref{eq:model-npred-calibration}. The columns represent the four different PSF event classes of \fermi. The color scale is shown in the right panel. More information on this analysis example can be found on \url{https://github.com/jolideco/jolideco-fermi-examples}. 
            }
            \label{fig:example-fermi-lat-non-calibrated}
            \script{example-fermi-lat-non-calibrated.py}
        \end{centering}
    \end{figure*}

    In this section we present an additional figure (Figure~\ref{fig:example-fermi-lat-non-calibrated} illustrating the deconvolution of \fermi data of \textit{RX J0852.0-4622}) and table (Table~\ref{tab:chandra_obs} describing the characteristics of the analyzed dataset).
    
    \label{sec:chandra-table}
    \begin{table}
    \centering
        \begin{tabular}{ c c c c c }
            Obs ID & Obs Date & Exposure & Rel. Exposure & Frac. Exposure \\
            \hline
             &  & $\mathrm{ks}$ &  & $\mathrm{\%}$ \\
            \hline
            8365 & 02/11/2007 & 21.0 & 1.0 & 6.1 \\
            6766 & 06/06/2006 & 19.7 & 0.9 & 5.8 \\
            9694 & 02/07/2008 & 19.2 & 0.9 & 5.6 \\
            15467 & 01/28/2013 & 19.1 & 0.9 & 5.6 \\
            14258 & 01/12/2012 & 19.0 & 0.9 & 5.6 \\
            13093 & 02/01/2011 & 19.0 & 0.9 & 5.6 \\
            11957 & 12/30/2009 & 18.4 & 0.9 & 5.4 \\
            6759 & 03/21/2006 & 17.9 & 0.9 & 5.2 \\
            17380 & 02/28/2015 & 17.7 & 0.8 & 5.2 \\
            26987 & 03/21/2023 & 14.4 & 0.7 & 4.2 \\
            25618 & 03/25/2022 & 14.4 & 0.7 & 4.2 \\
            22805 & 02/08/2020 & 14.3 & 0.7 & 4.2 \\
            21804 & 03/17/2019 & 14.3 & 0.7 & 4.2 \\
            20639 & 03/12/2018 & 14.3 & 0.7 & 4.2 \\
            19850 & 03/19/2017 & 14.3 & 0.7 & 4.2 \\
            18418 & 03/15/2016 & 14.3 & 0.7 & 4.2 \\
            24577 & 02/10/2021 & 13.8 & 0.7 & 4.0 \\
            17688 & 07/17/2015 & 9.6 & 0.5 & 2.8 \\
            16589 & 03/27/2014 & 9.6 & 0.5 & 2.8 \\
            6758 & 03/19/2006 & 8.1 & 0.4 & 2.4 \\
            10656 & 03/06/2009 & 7.8 & 0.4 & 2.3 \\
            6765 & 03/19/2006 & 7.6 & 0.4 & 2.2 \\
            10654 & 03/01/2009 & 7.3 & 0.3 & 2.1 \\
            10655 & 03/01/2009 & 6.8 & 0.3 & 2.0 \\
            \hline
        \end{tabular}
        \label{tab:chandra_obs}
        \script{chandra-obs-table.py}
        \caption{Observation list and details from the \chandra \textit{1E~0102.2–7219} example analysis in Section~\ref{sec:chandra-example}. The relative exposure columns list the exposure for each observation expressed as the ratio with the exposure of the reference observation with obs ID 8365. The fractional exposure column, gives the exposure of each observation as a fraction of the total exposure.}
    \end{table}

\end{document}